\documentclass[aps,prb,twocolumn,amsmath,amssymb,superscriptaddress]{revtex4-1}
\usepackage{graphicx,color,array}
\usepackage{subfigure}
\newcommand{\vheff}{\vec{h}_{\textrm{eff}}}
\newcommand{\vm}{\vec{m}}
\newcommand{\vx}{\vec{x}}
\newcommand{\vX}{\vec{X}}
\newcommand{\vV}{\vec{V}}
\newcommand{\Ms}{M_{\textrm{s}}}
\newcommand{\cN}{\mathcal{N}}
\newcommand{\cvP}{\vec{\mathcal{P}}}

\newcommand{\cE}{\mathcal{E}}

\begin{document}

\title{Propagation and control of nano-scale magnetic droplet
  solitons}

\date{\today}

\author{M. A. Hoefer}
 \email{mahoefer@ncsu.edu}
 \homepage{http://www4.ncsu.edu/~mahoefer/}
 \affiliation{Department of Mathematics, North Carolina State
 University, Raleigh, North Carolina 27695, USA}
\author{M. Sommacal}
 \email{msommac@ncsu.edu}
 \affiliation{Department of Mathematics, North Carolina State
 University, Raleigh, North Carolina 27695, USA}
\author{T. J. Silva}
 \email{silva@nist.gov}
 \affiliation{National Institute of Standards and Technology, Boulder,
   Colorado 80305, USA}

\begin{abstract}
  The propagation and controlled manipulation of strongly nonlinear,
  two-dimensional solitonic states in a thin, anisotropic ferromagnet
  are theoretically demonstrated.  It has been recently proposed that
  spin-polarized currents in a nanocontact device could be used to
  nucleate a stationary dissipative droplet soliton.  Here, an
  external magnetic field is introduced to accelerate and control the
  propagation of the soliton in a lossy medium.  Soliton perturbation
  theory corroborated by two-dimensional micromagnetic simulations
  predicts several intriguing physical effects, including the
  acceleration of a stationary soliton by a magnetic field gradient,
  the stabilization of a stationary droplet by a uniform control field
  in the absence of spin torque, and the ability to control the
  soliton's speed by use of a time-varying, spatially uniform external
  field.  Soliton propagation distances approach 10 $\mu$m in low loss
  media, suggesting that droplet solitons could be viable information
  carriers in future spintronic applications, analogous to optical
  solitons in fiber optic communications.
\end{abstract}

\maketitle

%
% Body Text
%

%%%%%%%%%%%%%%%%%%%%%%%%%%%%%%%%%%%%%%%%%%%%%%%%%%%%%%%%%%%%%%%%%%%%%%%%%%%%%%%%%%%%

\section{Introduction}
\label{sec:introduction}

Nanomagnetism holds great promise for future spin-based information
storage and processing technologies
\cite{bader_spintronics_2010,*lau_magnetic_2011}.  One enabling
physical effect is spin torque
\cite{slonczewski_current-driven_1996,*berger_1996}, which imparts
angular momentum from a spin-polarized current to a magnet.  Spin
torque forms the basis for tunable microwave nano-oscillators in
\emph{confined} nanopillar structures \cite{katine_device_2008} and
nanocontacts abutting an \emph{extended} ferromagnet
\cite{silva_developments_2008}.  Most nanopillar dynamics can be
reasonably described by single-domain modeling
\cite{berkov_spin-torque_2008} with the notable exception of
gyrotropic vortex motion \cite{yu_images_2011}.  In contrast,
nanocontacts enable the excitation of radiating
\cite{slonczewski_excitation_1999,bonetti_experimental_2010,demidov_direct_2010,*madamim._direct_2011}
and localized, coherently precessing, nonlinear wave states
\cite{bonetti_experimental_2010}.  The analysis of solitonic waves in
nanocontact systems has predominantly been limited to either the
weakly nonlinear regime at threshold \cite{slavin_spin_2005} or
complex micromagnetic simulations
\cite{berkov_magnetization_2007,*consolo_excitation_2007,berkov_spin-torque_2008}.

We recently proposed that a spin torque driven nanocontact could act
as a soliton creator in a uniaxial ferromagnet with sufficiently
strong perpendiular anistropy \cite{hoefer_theory_2010}.  The
resultant strongly nonlinear, coherently precessing state was termed a
dissipative droplet soliton, the locally driven/uniformly damped
cousin of the two-dimensional, nontopological droplet soliton
\cite{kosevich__1977,*kovalev_1979,*kosevich_magnetic_1990}.  Prior
numerical computations suggested that the conservative, stationary
droplet could be generalized to a propagating solution
\cite{piette_localized_1998}.  Small amplitude droplets were then
shown to propagate as approximate, Nonlinear Schr\"{o}dinger bright
solitons in [\onlinecite{ivanov_2001}].  The construction and
properties of a stable, two-parameter family of large amplitude
propagating droplet solutions in a lossless medium was undertaken in
[\onlinecite{hoefer_propagating_2011}].  However, a viable method to
accelerate solitons and understand their propagation in physically
realistic, damped media is lacking.

In this work, we use soliton perturbation theory to semi-analytically
demonstrate the feasibility of sustaining, moving, and controlling a
droplet soliton in a damped medium solely under the action of an
external magnetic field.  Modulation equations describing the
evolution of the soliton's speed and precessional frequency in the
presence of damping and a temporally/spatially varying external field
are studied and the results are corroborated by 2D micromagnetic
simulations.  We show that a stationary droplet can be accelerated by
a field gradient.  Once in motion, the soliton's speed can be
controlled by a spatially uniform, time-varying external field.  A
field gradient due to two nanowires can accelerate a soliton to
propagate approximately 10 $\mu$m in a low loss ferromagnet.
Stationary droplets of any allowable frequency can be created from a
sufficiently large, localized magnetic excitation, induced by a
nanocontact or otherwise, and then stabilized by a linear feedback
control field without the use of spin torque.  This represents a new
mechanism to study magnetic solitons without strong, spin torque
induced perturbations.  Analogous to optical solitons in fiber-optic
telecommunications \cite{mollenauer_solitons_2006}, these results show
that droplet solitons act as stable, controllable, particle-like,
precessing dipoles that exhibit intriguing nonlinear physics and hold
potential for spintronic applications.

The layout of this work is as follows.  First, we introduce the model
equations and then proceed with the finite dimensional reduction via
soliton perturbation theory.  The reduced system enables a thorough
analysis of droplet dynamics under the influence of damping and a
spatio-temporal magnetic field undertaken in the next section.  Two
control mechanisms, feedback control of a stationary droplet's
frequency and open loop control of a propagating droplet's speed, are
then introduced.  We conclude with some discussion and future outlook.

\section{Model}
\label{sec:model}

The model of magnetization dynamics we consider is the Landau-Lifshitz
equation \cite{berkov_spin-torque_2008}
\begin{equation}
  \label{eq:landaulifshitz}
  \begin{split}
     \frac{\partial \vm}{\partial t} = &- \vm
    \times \vheff - \alpha\,\vm \times (\vm \times \vheff ) \\
     \vheff = & \nabla^{2} \vm + (h_{0} + m_{z})\hat{z}, 
    % \frac{\partial \Theta}{\partial t} &= F - \alpha (G + h_0 \sin
    % \Theta), \\
    % \sin
    % \Theta \frac{\partial \Phi}{\partial t} &= G + h_0 \sin \Theta +
    % \alpha F, \\
    % F &= \frac{\nabla \cdot ( \sin^2 \Theta \nabla \Phi)}{\sin
    %   \Theta}, \\
    % G &= -\nabla^2 \Theta + \frac{1}{2} \sin 2 \Theta ( 1 + | \nabla
    % \Phi |^2 ),
  \end{split}
\end{equation}
describing a thin, two-dimensional, unbounded, damped ($\alpha > 0$ is
the damping parameter) ferromagnet.  The effective field incorporates
exchange $\nabla^2 \vm$, an external magnetic field $h_0(\vx,t)
\hat{z}$ pointing in the perpendicular direction normal to the film
plane, and perpendicular anisotropy $m_z \hat{z}$.  Crystalline
anisotropy, characterized by the anisotropy field $H_\textrm{k}$, is
assumed sufficient to overcome the local demagnetizing field so that
$H_\textrm{k} > M_\textrm{s}$.  Time, space, and fields are normalized
by scaled versions of the Larmor frequency $|\gamma| \mu_0 \Ms(Q -
1)$, exchange length $L_{\textrm{ex}}/\sqrt{Q-1}$, and saturation
magnetization $\Ms(Q-1)$, respectively, where $Q = H_{\textrm{k}}/\Ms
> 1$.  % is the
% quality factor measuring the relative strength of the anisotropy
% energy to magnetostatic energy.
We note that for the solitons studied here, the magnetostatic field is
approximately local for films with thickness much smaller than
$L_{\textrm{ex}}/\sqrt{Q-1}$ (see discussion in
\cite{hoefer_propagating_2011}).  For Co/Ni multilayer anisotropic
ferromagnets used in recent experiments
\cite{rippard_spin-transfer_2010,*mohseni_high-frequency_2011}, the
temporal scale and length scale are approximately 27 ps and 17 nm,
respectively, and $\alpha \approx 0.01$, $Q \approx 1.25$,
$M_\textrm{s} \approx 650$ kA/m.  References to dimensional results
use these parameter values.

In what follows, we assume that a localized excitation of large
amplitude has been nucleated by a spin torque nanocontact
\cite{hoefer_theory_2010} or some other means.  The rest of this work
is concerned with the manipulation of this structure in a lossy medium
by use of an external field.

When $h_{0} = \alpha = 0$, eq.~(\ref{eq:landaulifshitz}) admits the
conservation of total spin, momentum, and energy
\begin{align*}
  \cN &= \int (1 - \cos \Theta) d\vx, \\
  \cvP &= \int (\cos \Theta - 1) \nabla \Phi d\vx, \\
  \cE_0 &= \frac{1}{2} \int [ | \nabla \Theta |^2 + \sin^2 \Theta (1 + 
  |\nabla \Phi |^2)] d\vx,
\end{align*}
respectively, where all integrals are taken over the plane and
$\Theta$, $\Phi$ are the polar and azimuthal angles of the
magnetization, respectively.  Minimizing the energy subject to fixed
$\cN$ and $\cvP$ leads to a two-parameter family of localized,
precessing, stable traveling waves called propagating droplet solitons
parameterized by their velocity $\vV$ and frequency in the comoving
frame $\omega$ \cite{hoefer_propagating_2011}.  There is a bijective
map from $(\cN,\cvP)$ to the physical parameters $(\omega,\vV)$.
Droplet localization requires that the velocity and frequency of the
propagating droplet lie below the spin wave band, enforcing the
restriction
\cite{kosevich__1977,*kovalev_1979,*kosevich_magnetic_1990}
\begin{equation}
  \label{eq:3}
  \begin{split}
    \omega &+ |\vV|^2/4 < 1 , \quad \vV \ne 0, \\
    &0 < \omega < 1, \quad \vV = 0 .
  \end{split}
\end{equation}
We note that it is possible for moving droplets to exhibit negative
rest frequencies $\omega < 0$ \cite{hoefer_propagating_2011}.  Typical
droplet widths are of order one, hence are nanoscale excitations.
Stationary droplets with rest frequencies close to zero resemble
static circular bubbles, which received a great deal of attention in
the past \cite{leeuw_dynamic_1980}.  However, typical bubble sizes are
much larger.  With the inclusion of nonlocal magnetostatic fields,
Thiele [\onlinecite{thiele_theory_1970}] predicted that a static
bubble will be stable for a 5 nm thick film with a radius above 63
$\mu$m.  Thus, droplets can be viewed as smaller, dynamic
generalizations of the static bubble.

Allowing for weak damping ($\alpha \ll 1$) and a slowly varying
magnetic field ($|\nabla h_0 |, ~ |\partial_t h_{0}| \ll 1$), with no
restriction on the magnitude of $h_0$, causes the total spin,
momentum, and energy to evolve in time.  Through the map to
$(\omega,\vV)$, we can describe the droplet's adiabatic, particle-like
evolution by a time dependence of the droplet's velocity and rest
frequency, trajectories in the $V$-$\omega$ phase plane.

\section{Finite Dimensional Reduction}
\label{sec:low-dimens-dynam}

Using soliton perturbation theory (see, e.g.,
[\onlinecite{kivshar_dynamics_1989}]), we obtain the following finite
dimensional system of modulation equations describing the slow
modulation of the total spin and momentum:
\begin{subequations}
\label{eq:modeqs}
\begin{align}
  \label{eq:1}
  \frac{d \cN}{d t} = &- \alpha (\omega + h_0) \int \sin^2
  \Theta \,d\vx \\
  \nonumber
  &- \alpha \vV \cdot \int \sin^2 \Theta \nabla \Phi \, d \vx, \\
  \label{eq:2}
  \frac{d \cvP}{d t} = & \, - \nabla h_0 \mathcal{N} + \alpha ( \omega
  + h_0 ) \int \sin^2
  \Theta \nabla \Phi \, d\vx \\
  \nonumber & - \alpha \vV \cdot \int ( \nabla \Phi \sin^2 \Theta
  \nabla \Phi + \nabla \Theta \nabla \Theta)\, d\vx .
\end{align}
\end{subequations}
The energy $\cE = \cE_0 + \frac{1}{2} \int h_0 (1 - \cos \Theta) \,
d\vx$ is constrained to evolve according to
\begin{equation}
  \label{eq:8}
  \frac{d\mathcal{E}}{dt} = (\omega + h_0) \frac{d \mathcal{N}}{d t} +
  (\partial_t h_0 + \vV \cdot \nabla h_0) \mathcal{N} + \vV \cdot
  \frac{d\cvP}{dt} ;
\end{equation}
thus, it is sufficient to evolve eqs.~(\ref{eq:modeqs}) only.  The
integrals are evaluated with conservative droplets $(\Theta,\Phi)$ of
given total spin $\mathcal{N}(t)$ and momentum $\cvP(t)$ or,
equivalently, rest frequency $\omega(t)$ and velocity $\vV(t)$.  The
slowly varying field is evaluated along the soliton trajectory $h_0 =
h_0(\vX(t),t)$, where $d\vX/dt = \vV$.  Similar modulation equations
were derived for one-dimensional droplets in
[\onlinecite{baryakhtar_theory_1986,baryakhtar_soliton_1997,kosevich_magnetic_1998,*babich_relaxation_2001}]
and stationary, two-dimensional droplets ($\cvP \equiv 0$) in
[\onlinecite{baryakhtar_soliton_1997}].  Without loss of generality,
% Due to the rotational invariance
% of eq.~(\ref{eq:landaulifshitz}), 
we limit further discussion to
droplet motion in the $x$ direction, so that $\vV = (V,0)$ and $\cvP =
(\mathcal{P},0)$. % without loss of generality.

\begin{figure}
  \centering
  \subfigure{\includegraphics[scale=1]{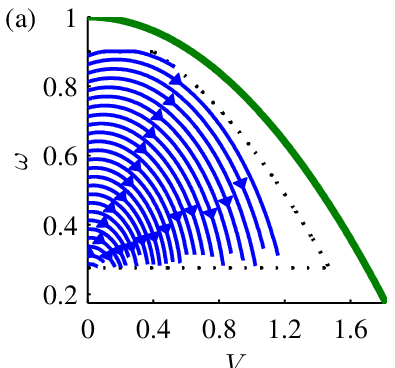}}
  \subfigure{\includegraphics[scale=1]{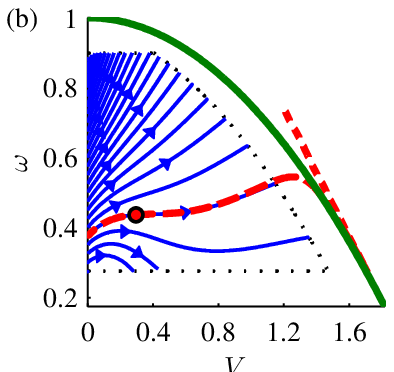}}\\[-2mm]
  \subfigure{\includegraphics[scale=1]{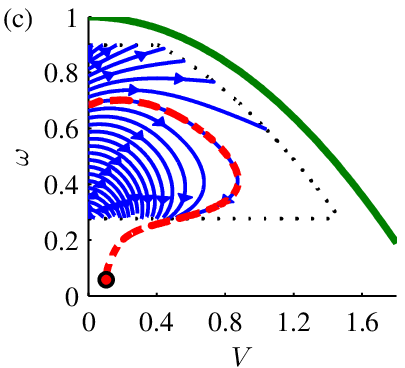}} 
  \subfigure{\includegraphics[scale=1]{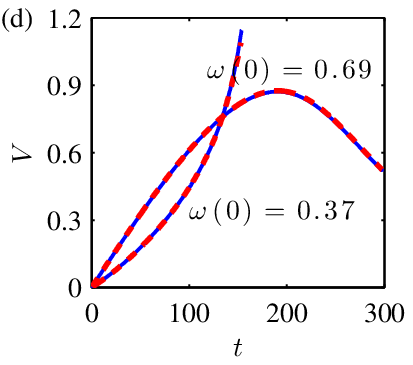}}\\[-2mm]
  \subfigure{\includegraphics[scale=1]{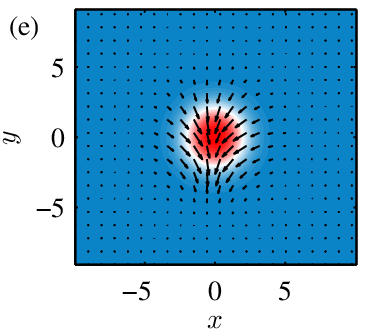}}
  \subfigure{\includegraphics[scale=1]{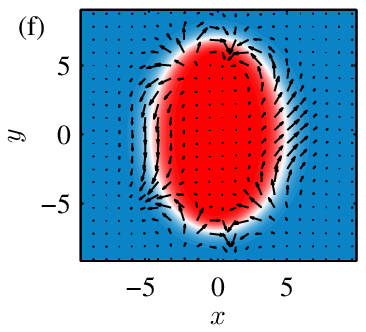}}
  \subfigure{\includegraphics[scale=1,clip=true,trim=78 -10 3 10]{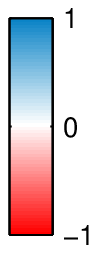}}
  \caption{(a,b,c) Droplet trajectories in the $V$-$\omega$ plane from
    numerical integration of the modulation equations
    (\ref{eq:modeqs}) (solid) and micromagnetics (dashed).  The dotted
    curves correspond to the edges of the precomputed droplet library
    and the thick, solid parabolic curve is the spin wave band.  (a)
    low loss $\alpha = 0.001$.  (b) $\alpha = 0.01$, positive bias
    field.  (c) $\alpha = 0.01$, negative bias field. 
    See text for further details.
    % with damping $\alpha = 0.001$ (a), $\alpha = 0.01$ (b,c), the
    % magnetic fields $h_0 = 0.5
    % - 0.005 x$ (a,b) and $h_0 = - 0.5 - 0.005 x$ (c).
    % The thick,
    % solid parabolic curves are the boundary of the inequality
    % (\ref{eq:3}), enclosing stable droplet soliton states.
    (d) Droplet velocity time dependence for the solid/dashed
    trajectories in (b,c).  (e,f) Droplet profiles from micromagnetics
    corresponding to the circles in (b,c), respectively.  The gray
    (color) scale represents the out-of-plane magnetization $m_z$ and
    the arrows represent the in-plane components.}
  \label{fig:1}
\end{figure}
In [\onlinecite{hoefer_propagating_2011}], we numerically computed a
library of propagating droplets using an iterative technique
\cite{library}.  These precomputed states are used here to numerically
solve the modulation equations \eqref{eq:modeqs} and to recover
$\omega(t)$ and $V(t)$.  The initial value problem for
eqs.~(\ref{eq:modeqs}) is numerically solved with initial parameters
$(V_0,\omega_0)$ chosen inside the precomputed library \cite{library}.
Interpolants mapping $(\mathcal{N},\mathcal{P})$ to $(V,\omega)$ and
vice-versa as well as for the integrals in (\ref{eq:modeqs}) are
generated from the droplet library.  We then numerically evolve
eqs.~(\ref{eq:modeqs}) in time using the interpolants.  We also
perform micromagnetic simulations of eq.~(\ref{eq:landaulifshitz}) by
use of a pseudospectral method \cite{hoefer_propagating_2011}.  To
recover the micromagnetic solution's speed and frequency, we compute
the center of mass $\int \vx (1 - \cos \Theta) \, d\vx/\mathcal{N}$
and the phase at the center of mass at each time step.  The velocity
and comoving frequency are then found by differentiation.  The rest
frequency is recovered by subtracting the local magnetic field
$h_0(\vX(t),t)$.

The modulation eqs.~(\ref{eq:modeqs}) represent a low dimensional
projection of the magnetodynamics enabling us to bring
finite-dimensional dynamical systems methods and control theory to
bear on the problem.  As we will demonstrate, the results from
modulation theory agree exceptionally well with micromagnetics.

\section{Droplet Soliton Dynamics}
\label{sec:dropl-solit-dynam}

The dynamics of eqs.~\eqref{eq:modeqs} depend on the magnitude of $|
\nabla h_0 |/\alpha$.  We consider each regime in turn.

\subsection{Negligible Damping:  $| \nabla h_0/\alpha | \gg 1$}
\label{sec:negligible-damping:-}

When $| \nabla h_0/\alpha | \gg 1$, the total spin is approximately
conserved and the momentum varies.  The $V$-$\omega$ phase plane for
$h_0 = 0.5-0.005x$ and $\alpha = 0.001$ pictured in
Fig.~\ref{fig:1}(a) closely resembles trajectories of constant total
spin, $\mathcal{N} = \mathcal{N}_0$ or constant energy depicted in
[\onlinecite{hoefer_propagating_2011}].  These low damping dynamics
can be approximated by setting $\alpha = 0$.  Then eqs.~(\ref{eq:1})
and (\ref{eq:2}) become Newton's law 
\begin{equation*}
  d \cvP/dt = - \mathcal{N}_0
  \nabla h_0(\vX,t), \quad d \vX/dt = \vV ,
\end{equation*}
for the motion of the soliton center $\vX(t)$ subject to the potential
$\mathcal{N}_0 h_0$.  Thus, engineering the magnetic field in an
appropriate way allows one to control the motion of the particle-like
soliton.  Note that the effective mass
\begin{equation}
  \label{eq:6}
  m_\mathrm{eff}(\omega,V) = \frac{\mathcal{P}(\omega,V)}{V},
\end{equation}
depends on the soliton speed and frequency.  The stationary droplet is
accelerated while the rest frequency decreases, representing a
transfer of the effective potential energy stored in the precessional
motion $\omega$ to effective kinetic energy of translational motion
$V$.

\subsection{Comparable Damping and Field Gradient: $| \nabla h_0/\alpha | =
\mathcal{O}(1)$}
\label{sec:comp-damp-field}

For the balance $| \nabla h_0 /\alpha | = \mathcal{O}(1)$, different
dynamics occur.  Figure \ref{fig:1}(b) depicts trajectories for the
same field as in \ref{fig:1}(a) but with and order of magnitude larger
damping $\alpha = 0.01$.  The dashed curve depicts the trajectory from
micromagnetics with $\omega(0) = 0.37$.  The circle on this trajectory
corresponds to the droplet shown in Fig.~\ref{fig:1}(e).  The droplet
is accelerated and is accompanied by an amplitude decrease until it
devolves into a linear spin wave upon reaching the band edge
(eq.~\eqref{eq:3}).  The micromagnetic simulation closely matches the
adiabatic theory until the band edge is reached and the solution
amplitude is very small, after which the droplet ansatz is no longer
valid.  A plot of $V(t)$ is shown in Fig.~\ref{fig:1}(d).

During the course of evolution, a droplet can experience deceleration,
as in Fig.~\ref{fig:1}(c) with $\alpha = 0.01$ and a negative bias
field, $h_0 = -0.5 - 0.005x$.  This behavior is reminiscent of Bloch
oscillations predicted for one-dimensional droplets in
[\onlinecite{kosevich_magnetic_1998,*babich_relaxation_2001}].  In the
one-dimensional (1D) case, the soliton oscillates under a constant
force.  We have not observed such behavior in our micromagnetic
simulations or in the modulation theory.  Instead, simulations reveal
the formation of local, topological structure including
vortex/anti-vortex pairs and what appears to be switching of the
magnetization to the inverted, $\Theta \equiv \pi$ state,
corresponding to $(V,\omega) = (0,0)$ shown in Fig.~\ref{fig:1}(f).
Previous work on solitons in \emph{isotropic} ferromagnets reveal the
existence of nontopological droplets bifurcating into co-propagating
vortex, anti-vortex pairs when a critical momentum is reached
\cite{cooper_1998}.  Approximate vortex, anti-vortex pairs were
studied in the anisotropic case numerically
\cite{piette_localized_1998} but have not been observed in our
micromagnetic studies here.

\begin{figure}
  \centering
  \includegraphics[scale=1]{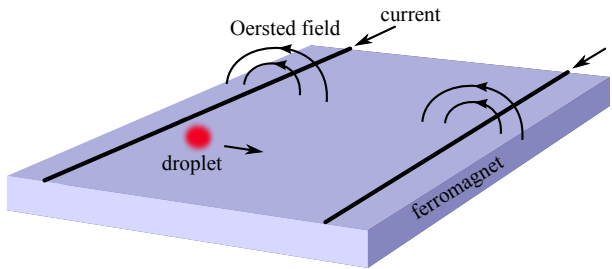}
  \caption{Droplet acceleration by current flowing through two
    nanowires.}
  \label{fig:two_wires}
\end{figure}
\begin{figure*}
  \centering
  \subfigure{\includegraphics[scale=1]{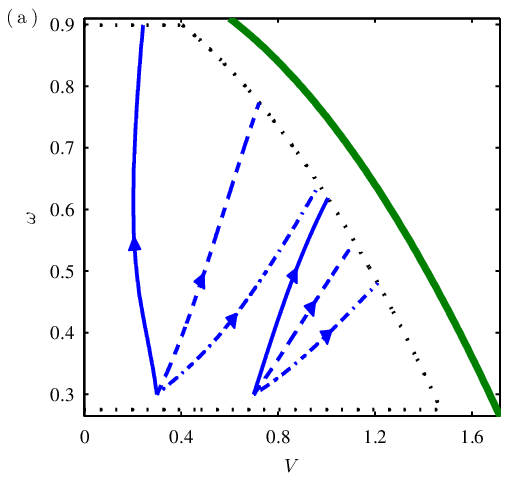}}
  \subfigure{\includegraphics[scale=1]{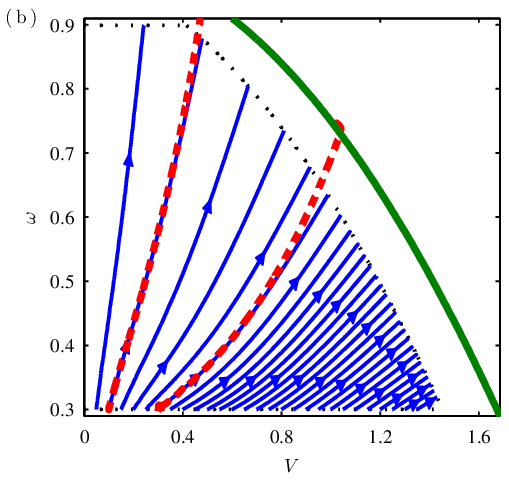}}
  \subfigure{\includegraphics[scale=1]{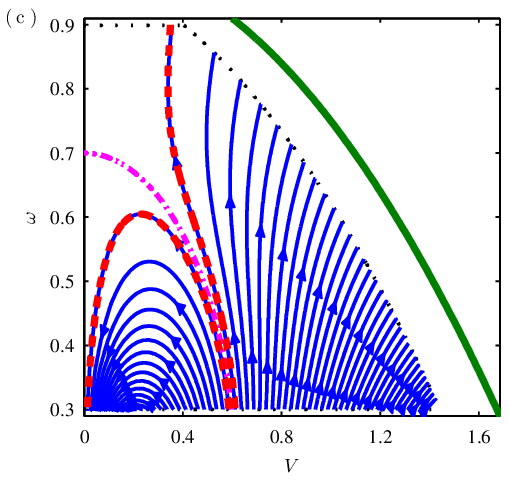}}
  \caption{Droplet trajectories with uniform, static magnetic field
    and $\alpha = 0.01$.  (a) Modulation solution showing acceleration
    or deceleration of a propagating droplet due to damping and $h_0 =
    -0.4$ (solid), $h_0 = 0$ (dashed), and $h_0 = 0.9$ (dash-dotted).
    (b) $h_0 = 1$.  (c) $h_0 = -0.7$.  In (b) and (c), the dashed
    curves are micromagnetic simulations.  In (c), the dash-dotted
    curve is the separatrix between the switched state $(V,\omega) =
    (0,0)$ and spin wave states.}
 \label{fig:2}
\end{figure*}
In addition to the dynamics of Fig.~\ref{fig:1} for constant field
gradient, we have also simulated droplet acceleration due to the field
generated by two current-carrying nanowires in the plane of the film
with current in the same direction and a stationary droplet nucleated
in between them (Fig.~\ref{fig:two_wires}).  The current induced
Oersted fields lead to a negative magnetic field gradient that
accelerates the droplet.  The low dimensional modulation system,
eq.~(\ref{eq:modeqs}), enables a detailed investigation of parameter
space that would, using micromagnetic simulations, be prohibitive to
explore.  The simulations incorporate the Oersted field due to two
infinite wires with 150 nm diameters and varying separation.
Stationary droplets with varying frequencies are assumed to be
nucleated 300 nm from the center of the left wire and allowed to
propagate until either reaching the second wire or the soliton center
of mass attains the value $m_z = 0.5$.  For the moderate damping case
$\alpha = 0.01$, the droplet is predicted to travel up to 3 $\mu$m in
about 30 ns for 15 mA current in each wire with a 3.3 $\mu$m wire
separation.  Top speeds can approach 600 m/s.  In the low-loss case
$\alpha = 0.001$, the droplet can propagate about 10 $\mu$m in 70 ns
for a 10 mA current with a 10.3 $\mu$m wire separation.

\subsection{Negligible Field Gradient:  $ |\nabla h_0|/\alpha \ll 1$}
\label{sec:negl-field-grad}

The remaining regime, when $|\nabla h_0|/\alpha \ll 1$, is now
investigated.  We focus on the case of a uniform and static magnetic
field, assuming that a propagating droplet has been created.  In
Fig.~\ref{fig:2}, solution of the modulation system (\ref{eq:modeqs})
reveals the \emph{acceleration} of a propagating droplet due to
damping when the magnetic field is zero or positive.  When the field
is sufficiently negative, the droplet can experience deceleration and
then acceleration as its amplitude decays.  This counter-intuitive
droplet acceleration due to damping was predicted for 1D droplets in
the absence of a magnetic field \cite{baryakhtar_theory_1986}.  We can
understand this behavior in terms of the droplet's effective mass
\eqref{eq:6}.  From eq.~\eqref{eq:6} we have $\dot{P} =
\dot{m}_\mathrm{eff} V + m_\mathrm{eff} \dot{V}$ where $\dot{\ }$
denotes time differentiation.  In the absence of a field gradient,
eq.~\eqref{eq:2} implies a decrease in momentum $\dot{P} < 0$.  Then a
droplet can be accelerated ($\dot{V} > 0$) if 
\begin{equation}
  \label{eq:9}
  \dot{m}_\mathrm{eff} < \dot{P}/V < 0 .
\end{equation}
In other words, the droplet is accelerated in the presence of damping
because the effective mass is decreasing at a sufficiently fast rate.

As shown in Fig.~\ref{fig:2}(b), inequality \eqref{eq:9} holds for
$\omega > 0.3$ and positive fields.  When $-1 < h_0 < 0$, there are
some droplets that exhibit deceleration.  In this case, the magnet
undergoes a complete reversal to the $(V,\omega) = (0,0)$ state for
initial droplets with parameters lying below a separatrix (see
Fig.~\ref{fig:2}(c)) which we term the \emph{switching separatrix}.
The switching separatrix corresponds to the stable manifold of the
fixed point $(V,\omega) = (0,-h_0)$.  Linearization of
eqs.~(\ref{eq:modeqs}) around this fixed point results in the
eigenvalues
\begin{equation}
  \label{eq:5}
  \left ( \frac{-\int \sin^2
      \Theta \, d\vx}{\partial_\omega \mathcal{N}}, \quad \frac{- \int
    \Theta_x^2 \, d\vx}{\partial_V \mathcal{P}} \right ),
\end{equation}
evaluated at $(V,\omega) = (0,-h_0)$.  Since $\partial_\omega
\mathcal{N} < 0$ \cite{kosevich_magnetic_1990} and $\partial_V
\mathcal{P} > 0$ for stationary droplets, the fixed point is a saddle.
The switching separatrix from modulation theory accurately resolves
the micromagnetic dynamics as evidenced by the close agreement in
Fig.~\ref{fig:2}(c) for trajectories starting very close to the
separatrix.  The switching separatrix for negative bias field leads to
the differing phase plane trajectories in Figs.~\ref{fig:1}(b) and
(c).  Physically, this analysis reveals that there is a synergy
between damping and the bias field.  A bias above $-\omega$ leads to a
decrease of the soliton amplitude whereas a bias below causes an
increase in soliton amplitude.  This suggests a mechanism for
stabilizing a stationary droplet at the saddle point $(V,\omega) =
(0,-h_0)$ by dynamically changing the bias field with feedback
control.

\section{Droplet Soliton Control}
\label{sec:dropl-solit-contr}

\subsection{Stationary Droplet Stabilization}
\label{sec:stat-dropl-stab-1}

The nucleation of a stationary droplet by a spin torque driven
nanocontact has been theoretically demonstrated
\cite{hoefer_theory_2010}.  The droplet, which would otherwise decay
due to damping, is sustained by a balance between localized driving
and uniform damping.  However, the current-induced Oersted field
strongly perturbs the stationary droplet from its ideal, symmetric
structure, leading to phase variations and potentially a drift
instability whereby the droplet is ejected from the nanocontact
\cite{hoefer_theory_2010}.  Furthermore, canting of the polarization
layer is required to obtain an ac electrical signal via giant
magnetoresistance leading to symmetry breaking of the spin torque term
and further complexity.  We propose a simple alternative stabilizing
mechanism that avoids these difficulties: a closed-loop, spatially
uniform control field.  The stationary droplet saddle point
$(V,\omega) = (0,\omega_*)$ for $h_0 = - \omega_*$ and $0 < \omega_* <
1$ is altered to an attractor by introducing the linear feedback
control
\begin{equation}
  \label{eq:4}
  h_0(t) = -\omega_* + G \Omega(t),
\end{equation}
with gain $G$, bias $-\omega_*$, and the total, measured system
frequency $\Omega(t) = \omega(t) + h_0(t)$.  Feedback implementation
could be realized by use of a small amplitude (sub-threshold) dc
current applied to a trilayer nanocontact with a canted fixed layer,
resulting in a measurement of $\Omega$ with negligible spin torque and
Oersted field effects.  Another possibility is direct imaging of the
dynamics.

The goal is to drive $\Omega$ to zero.  With such a control law,
linearization of eq.~(\ref{eq:modeqs}) around the fixed point results
in the eigenvalues
\begin{equation}
  \label{eq:1}
  \left ( \frac{\int \sin^2
      \Theta \, d\vx}{(G - 1) \partial_\omega \mathcal{N}}, \quad \frac{- \int
      \Theta_x^2 \, d\vx}{\partial_V \mathcal{P}} \right ) .
\end{equation}
Since $\partial_\omega \mathcal{N} < 0$ and $\partial_V \mathcal{P} >
0$ \cite{hoefer_propagating_2011}, when $G > 1$ both eigenvalues are
negative and the fixed point is linearly stable.  This proves linear
stability.  However, if the initial system frequency is too close to
1, one may apply the feedback field (\ref{eq:4}) with $h_0 < -1$
leading to spontaneous reversal of the magnetic film.  To avoid this
scenario and stabilize the droplet, the gain is restricted to
\begin{equation*}
  G > \max\left ( 1, 1 + \omega(0) - \omega_* \right
  ) .
\end{equation*}
The gain $G$ determines the relaxation rate to the fixed point, with
larger values leading to slower relaxation.  For excitations with
$\omega(0) - \omega_* > 0$, $G$ cannot be very close to 1.  Thus,
there is a trade-off between the relaxation time and the stability of
the ferromagnet.

\begin{figure}
  \centering
  \subfigure{\includegraphics[scale=1.3]{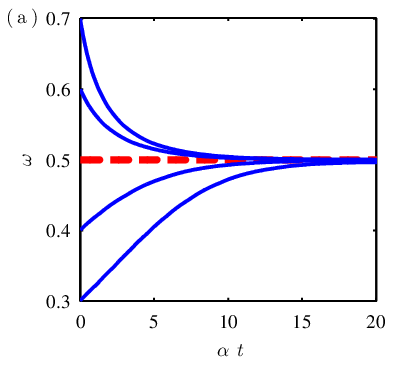}}
  \subfigure{\includegraphics[scale=1.3]{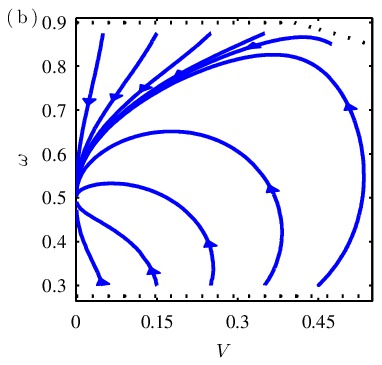}}
  \caption{Stabilization of a stationary droplet with $\omega_* = 0.5$
    by the linear feedback control law (\ref{eq:4}) with $G = 2$.  (a)
    Stationary droplet relaxation.  (b) Trajectories in the
    $V$-$\omega$ plane.}
  \label{fig:3}
\end{figure}
We observe that the fixed point is a global attractor of the
modulation equations (\ref{eq:modeqs}) in Fig.~\ref{fig:3}.  Figure
\ref{fig:3}(a) shows the relaxation of the frequency to $\omega_* =
0.5$ for stationary droplets.  If $V(0) = 0$, and the control law
(\ref{eq:4}) is assumed, then eq.~(3) simplifies to
\begin{equation}
  \label{eq:7}
  \frac{d \omega}{dt} =  \alpha\frac{(\omega_* -
    \omega)}{(1 - G) \partial_\omega \mathcal{N}}  \int
  \sin^2 \Theta \, d\vx ,
\end{equation}
with $V(t) \equiv 0$.  Since $\partial_\omega \mathcal{N} < 0$ and $G
> 1$, we observe that $\omega = \omega_*$ is a global attractor for
fixed $V(t) = 0$ with $\omega(t)$ relaxing monotonically to
$\omega_*$.

When $V(0) \ne 0$, the initial state exhibits phase variations leading
to droplet propagation.  However, as Fig.~\ref{fig:3}(b) shows,
$(\omega,V) = (\omega_*,0)$ is still an attractor so the phase
perturbations decay in time; hence, any drift instability, such as
that observed in the case of a nanocontact system
\cite{hoefer_theory_2010}, has been removed.

\begin{figure}
  \centering
  \includegraphics[scale=1]{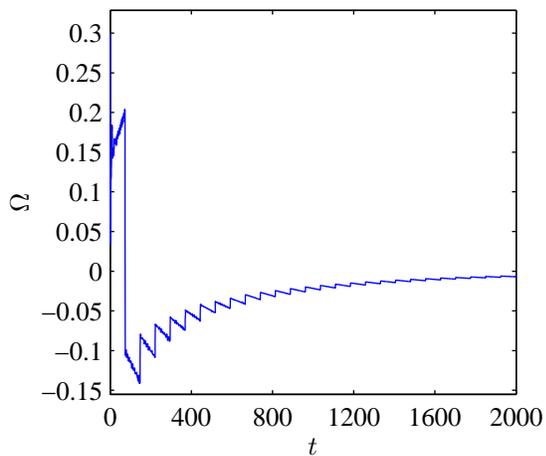}
  \caption{Micromagnetic simulation giving the spatially averaged
    magnetization frequency due to the control field (\ref{eq:4}),
    $\omega_* = 0.4$, $G = 1.5$, $\alpha = 0.01$, and an update period
    of 74.  The droplet is locked when $\Omega = 0$.}
  \label{fig:control_micromagnetics}
\end{figure}
We have also performed micromagnetic simulations of
eq.~(\ref{eq:landaulifshitz}) incorporating the feedback control law
(\ref{eq:4}).  We begin the computation with an asymmetric, localized
initial condition
\begin{equation}
  \label{eq:2}
  \Theta(x,y,0) = A e^{-(x/w_x)^2-(y/w_y)^2} , \quad \Phi(x,y,0) = 0 ,
\end{equation}
where $\vm = (\cos \Phi \sin \Theta, \sin \Phi \sin \Theta, \cos
\Theta)$ and $A = 2.7$, $w_x = 2.3$, $w_y = 3$.
Equation~(\ref{eq:landaulifshitz}) is evolved with the spatially
uniform field (\ref{eq:4}) where $\omega_* = 0.4$, $\alpha = 0.01$, $G
= 1.5$.  We ``measure'' $\Omega(t)$ by averaging the in-plane
magnetization orientation over the unit disk to obtain an average
phase $\overline{\Phi}(t)$.  This models the nanocontact measurement
technique suggested; then, $\Omega(t) =
\frac{d\overline{\Phi}}{dt}(t)$.  We perform computations with
differing bandwidths or update times so that $h_0(t)$ is updated
instantaneously and periodically.  Figure
\ref{fig:control_micromagnetics} depicts the evolution of the total
frequency $\Omega(t)$ with a field update period of 74 time units,
which translates to 2 ns or a 500 MHz bandwidth for perpendicular
magnets used in recent experiments
\cite{rippard_spin-transfer_2010,*mohseni_high-frequency_2011}.
Because damping drives the dynamics, we expect that the operable
control bandwidth is approximately $\alpha |\gamma|\mu_0 M_\textrm{s}
(Q - 1)$, corresponding to about 400 MHz.  This suggests that any
sufficiently large, localized excitation created by spin torque or
other means can be deformed into a stationary droplet of choice solely
with the use of an external, spatially uniform magnetic field.
Physically, this stabilization mechanism balances the switching of a
droplet by a sufficiently negative bias field and the decay of a
droplet via damping.

% \section{Extra stuff}
% \label{sec:extra-stuff}

% Related system: propagating, weakly nonlinear 2D, backward volume
% magnetostatic spin wave soliton packets excited by pulsed pumping of
% an external magnetic field observed in \cite{serga_parametric_2005}
% with diameters on the order of a millimeter.

\subsection{Droplet Speed Control}
\label{sec:dropl-speed-contr}

\begin{figure}[h]
 \centering
 \subfigure{\includegraphics[scale=1.5]{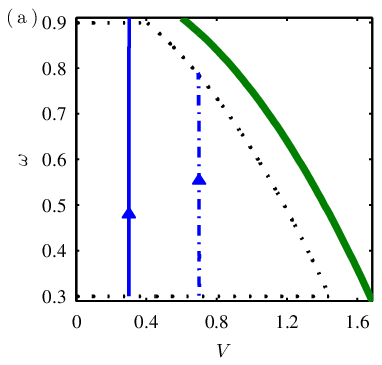}}
 \subfigure{\includegraphics[scale=1.5]{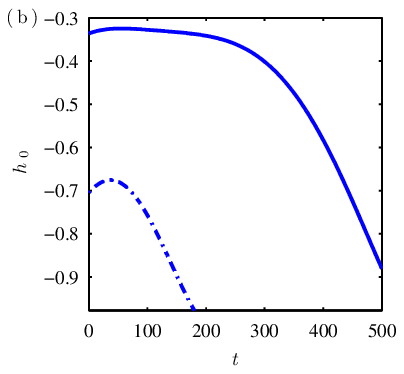}}
 \caption{Open-loop control of droplet speed.  (a) Two trajectories
   (solid, dash-dotted) in the $V$-$\omega$ plane from micromagnetic
   simulations with (b) the corresponding control field (solid,
   dash-dotted, respectively) determined from the modulation equations
   (3).}
 \label{fig:4}
\end{figure}
In the presence of damping and a constant magnetic field, a
propagating droplet is either accelerated or decelerated (recall
Fig. 2).  A time-varying, spatially uniform open-loop control field
can be used to stabilize the droplet's speed.  For this, we implement
an optimization strategy for the modulation equations
(\ref{eq:modeqs}) by stepping forward in time and determining the
field $h_0(t)$ that enforces the constraint of constant speed.  The
resulting field profile is fit to a quintic polynomial (see
Fig.~\ref{fig:4}(b)) and used in a micromagnetic simulation of
eq.~(\ref{eq:landaulifshitz}) resulting in the trajectories shown in
Fig.~\ref{fig:4}(a). Remarkably, the control field from modulation
theory leads to accurate control of the droplet's speed in the full
micromagnetic simulation.

\section{Conclusion}
\label{sec:conclusion}

We have demonstrated that droplets propagating in a realistic, damped
ferromagnet can be sustained, accelerated, and controlled by use of
only an external magnetic field.  Combining a nanocontact system
\cite{hoefer_theory_2010} with the ideas presented here provides a
framework to create and control moving droplet solitons in a
ferromagnet.  Their robustness and controllability hold promise for
future spintronic applications.

% \bibliography{damped_droplet_prl}
% \bibliographystyle{apsrev4-1}
%

\end{document}